\documentclass[runningheads]{llncs}
\usepackage[T1]{fontenc}
\usepackage{graphicx}
\usepackage{algorithm}
\usepackage{algpseudocode}
\usepackage{listings}
\usepackage{hyperref}
\usepackage{pgfplots}
\usepackage{tikz}
\usepackage{tabularx}
\begin{document}
\title{Divide, Conquer and Verify: Improving Symbolic Execution Performance}
%
%
\author{Christopher Scherb\inst{1} \and Luc Heitz\inst{1} \and
Hermann Grieder\inst{1} \and Olivier Mattmann\inst{1}}

\authorrunning{C. Scherb et al.}
%
\institute{University of Applied Sciences and Arts, Northwestern Switzerland
\email{christopher.scherb@fhnw.ch, luc.heitz@fhnw.ch, hermann.grieder@fhnw.ch, olivier.mattmann@fhnw.ch}\\
}
\maketitle              
\begin{abstract}
Symbolic Execution is a formal method that can be used to verify the behavior of computer programs and detect software vulnerabilities. Compared to other testing methods such as fuzzing, Symbolic Execution has the advantage of providing formal guarantees about the program. However, despite advances in performance in recent years, Symbolic Execution is too slow to be applied to real-world software. This is primarily caused by the \emph{path explosion problem} as well as by the computational complexity of SMT solving. In this paper, we present a divide-and-conquer approach for symbolic execution by executing individual slices and later combining the side effects. This way, the overall problem size is kept small, reducing the impact of computational complexity on large problems.

\keywords{Symbolic Execution  \and Formal Methods \and Formal Verification \and Program Verification}
\end{abstract}

\section{Introduction}
Since the dawn of the computer age, the complexity of software and hardware has continuously increased. As soon as new development tools and methods become available, developers push them to their limits. While this accelerates the development of new features, tools, and software, it also increases the risk of creating bugs and vulnerabilities. Mitigating these bugs and vulnerabilities is one of the primary challenges in modern software development. Various approaches have been developed to avoid bugs, ranging from testing individual components for their correct functionality using unit tests and fuzzing tests~\cite{zhu2022fuzzing} for detecting implementation errors and vulnerabilities, to employing formal methods to prove software correctness.

Formal methods such as Symbolic Execution can be used to detect a wide range of bugs, such as memory corruption, but they can also verify the functionality and business logic of software~\cite{cadar2013symbolic}. Therefore, Symbolic Execution is a very powerful tool in software verification. The advantages of formal methods compared to fuzzing and unit testing lie in the guarantees these methods provide.

However, the significant disadvantage of all formal methods is typically their computational complexity. Formal methods generally follow a principle of checking if any possible path violates the given correctness conditions, implying that all potential paths need to be evaluated. This leads to the path explosion problem, as the number of possible paths to check increases exponentially with the problem's size. Furthermore, testing if a path is feasible requires SMT solving, which is at least NP-hard depending on the theories used.

Therefore, Symbolic Execution, like all formal methods, is highly effective but often impractical for most real-world problems. These problems are too large and cannot be solved within a reasonable timeframe or memory usage. As there are no algorithms with lesser computational complexity to expedite formal methods, the solution to larger problems lies in dividing them into smaller subproblems and subsequently recombining them, following the so-called divide-and-conquer method. By limiting the size of the smaller subproblems, it is possible to solve and later recombine them, even with NP-hard complexity.

However, finding efficient methods for dividing and recombining problems is challenging, especially when subproblems can have side effects and are in complex relations with each other. This complexity is evident in the context of computer programs, where each slice is in a complex relation with the rest of the program.

In this paper, we propose a method for Symbolic Execution that splits computer programs into smaller subprograms and creates abstract representations of each subproblem. Later, we recombine the subproblems to solve the entire task. For slicing a computer program, we leverage its natural structure, such as functions and loops, which are typically the most problematic structures for Symbolic Execution.

Our approach involves slicing computer programs by functions, although other types of slicing are also possible. We execute the slices symbolically while creating a symbolic summary. There are existing approaches to creating summaries of functions, such as by assuming a function will always return either a fixed concrete value or a symbolic value independent of the parameters. Some approaches execute only the relevant path for the given precondition to save execution time. If a different precondition for the function call is not found upon the second invocation of the same function, the previous result will be reused. Therefore, a function only needs to be re-executed if a different precondition is found for another invocation.

Our approach creates generalized symbolic summaries. The idea is to symbolically execute a function to find all possible paths and to define symbolic and concrete return values for all possible preconditions of the function. This creates a lookup table that indicates the expected results for each parameter. Both concrete and symbolic values are possible.

We evaluate our approach using various test programs. We created a dataset that contains specific computer programs, which are challenging for classic symbolic execution, to highlight our improvements.

The rest of the paper is structured as follows: In section~\ref{sec:RelatedWork}, we present relevant related work. In section~\ref{sec:SmartSummaries}, we explain how we generate and apply our symbolic summaries. The evaluation of our approach can be found in section~\ref{sec:Eval}. We conclude our paper in section~\ref{sec:Conclusion}.

\section{Background \& Related Work}
\label{sec:RelatedWork}

In this section we present the relevant related work and similar approaches to function wise program slicing for symbolic execution. At the end of this section we will present the search gap and explain how our approach differs from the existing approaches. For our literature review we searched in Google Scholar for \emph{Symbolic Execution}, \emph{Function Summaries for Symbolic Execution}, \emph{Function Summaries}, \emph{Slicing in Symbolic Execution}, \emph{Compositional Symbolic Execution} and \emph{Symbolic Summaries}. 
\subsection{Background}

\paragraph{Symbolic Execution} is a powerful software testing technique that operates on program variables as symbolic values instead of actual data. The technique generates symbolic expressions that represent the behavior of a program over all possible execution paths~\cite{king1976symbolic}. 

Symbolic Execution starts by assigning symbolic values, rather than actual data values, to the input variables of a program. These symbolic values denote all possible values that the input can take. It then runs the program with these symbolic values as inputs, in essence simulating the program's execution across all possible input values simultaneously.

As the symbolic execution progresses and encounters various operations and conditionals (like if-else statements), it modifies the symbolic values according to the operation encountered. For example, if it encounters an addition operation, it will generate a symbolic expression that corresponds to the sum of the two input variables. Similarly, for a conditional statement, symbolic execution will consider both branches, leading to two separate execution paths.

In this process, each path in the program is associated with a 'path condition', a logical formula over the input variables, which captures the conditions under which the path would be taken. For instance, for an if-else statement, there would be two paths: one path where the condition in the 'if' clause is true, and another where it is false. Each of these paths would have corresponding path conditions.
Symbolic execution maintains a 'symbolic state', capturing the current values of variables as symbolic expressions and the path condition for reaching the current execution point. When it encounters a branch, symbolic execution 'forks' the state into separate states for each branch.
After executing all paths, or when execution is halted due to practical constraints, the resulting symbolic expressions and path conditions can be used to generate test inputs that will execute each path, find potential bugs or security vulnerabilities, or prove properties of the program.

\paragraph{Static Symbolic Execution} is an offshoot of symbolic execution that focuses on analyzing a program without actually executing it. This static analysis allows for a comprehensive examination of all possible execution paths and conditions in a program, making it valuable for finding vulnerabilities, bugs, or violations of coding guidelines. However, its downside is the 'path explosion' problem, due to which it can become computationally infeasible for larger programs~\cite{cadar2013symbolic}.

\paragraph{Dynamic Symbolic Execution}, also known as concolic testing, combines concrete and symbolic execution~\cite{irlbeck2015deconstructing}. It operates by running the program with a concrete input while simultaneously maintaining symbolic expressions for the executed program paths. This dual nature helps mitigate the 'path explosion' problem and provides a more realistic view of program behavior. However, this also comes with the trade-off of potentially missing some paths due to the dependence on concrete executions~\cite{sen2005cute}.


\subsection{Related Work}
\label{sec:RelWork}
Over the long time symbolic execution exists, many approaches have been taken to enhance the speed of symbolic execution~\cite{baldoni2018survey}. 
The main idea of most optimization approaches is to tackle the path explosion problem.
The approaches range from probabilistic path selection to veritesting. 

\paragraph{Probabilistic path selection} is a strategy used in symbolic execution to tackle the path explosion problem, which arises due to the exponential growth of potential execution paths in a program. As it's often impossible to explore every single path within a reasonable time frame, particularly for large and complex programs, a path selection strategy is necessary to decide which paths to explore first or prioritize~\cite{geldenhuys2012probabilistic}.
The probabilistic path selection strategy is one such approach, where each unexplored path is assigned a probability, and the path to be explored next is chosen based on these probabilities. The way the probabilities are assigned can be determined by various factors. For example, one might assign higher probabilities to paths that have been less thoroughly explored or paths that appear to be leading towards potentially interesting behavior~\cite{burnim2008heuristics}.

\paragraph{Veritesting}~\cite{avgerinos2014enhancing} is an extension of symbolic execution aimed at mitigating the 'path explosion' problem, which is a common challenge in symbolic execution. Traditional symbolic execution suffers from a 'path explosion' problem as it treats each branch in a program as a separate path, which can lead to an exponential number of paths in the worst case, making it computationally expensive or even infeasible for large programs.
Veritesting mitigates this problem by attempting to merge multiple execution paths into a single symbolic formula. It does this by employing static analysis techniques to identify portions of the code (within a method or a loop, for example) where paths can be merged. For these identified regions, instead of creating a new path for each branch, veritesting creates a single formula that encapsulates all the paths. This approach significantly reduces the number of paths to be considered, thereby greatly improving the scalability of symbolic execution~\cite{avgerinos2014enhancing}.

Other approaches address the problem size by decomposing computer programs into smaller parts. A smaller problem size is easier and more efficient to execute. 
For example, the standard library usually is not executed, but mock-functions are used instead of the actual function calls. This way, also side effects like I/O can be simulated. A key property of the mock-functions is, to simulate the behavior of the actual function, but be faster. For example, the \emph{C Standard Library} function \texttt{strlen}, which takes a string and returns the length can be simulated by a function which returns a symbolic value. However, it is also possible to make the summary more precise by letting the symbolic solver try to search for a string position which can only be zero and assume this as maximum length~\cite{ramos2023toward}. These kind of summaries are reducing the search space for the symbolic execution engine by limiting the number of possible successors. However, creating very specialized function summaries is hard to automatize. 
\paragraph{Chopped symbolic execution} is a method that aims to address the path explosion problem and increase the efficiency of traditional symbolic execution. This method introduces a chopping criterion to symbolically execute only a relevant part of the program, thereby ignoring parts that don't affect the outcome of interest. The criterion is based on a user-specified property of the code or a set of variables the user is interested in~\cite{trabish2018chopped}.

\paragraph{Compositional Dynamic Test Generation} by Godefroid is a technique used in software testing to enhance the efficiency and scalability of dynamic symbolic execution~\cite{godefroid2007compositional}. The central idea behind compositional test generation is to separate the program into different components (for instance, functions or methods), symbolically execute each component separately, and reuse the results for subsequent tests.
This approach addresses the path explosion problem inherent in traditional symbolic execution, where the number of possible execution paths increases exponentially with program size, making exhaustive exploration infeasible for large programs. In Compositional Dynamic Test Generation, by breaking down the program into smaller components and reusing results, a larger portion of the code-base can be effectively covered with fewer computational resources.
For each component, Compositional Dynamic Test Generation symbolically executes the component, summarizing the effects of the component on its inputs and outputs. When a component is encountered during the execution of the larger program, instead of re-executing the component the summary computed earlier is used. Function are computed by successive iterations, only one path at the time~\cite{anand2008demand}. After the execution a path, the pre- and the post-conditions are extracted. If the same precondition is hit again, the cached post-condition will be applied. 

\paragraph{Fine-Grained Summaries}  is a concept introduced by Yude Lin et al.~\cite{lin2015compositional} which takes this idea further by focusing on the level of granularity at which the summaries are created. Summaries are representations of the effects that a certain part of the program (like a function or a method) has on its inputs and outputs. By generating summaries at a finer granularity—meaning at a more detailed level, such as individual lines of code or smaller blocks of code, instead of larger components like functions or methods—we can get more precise representations of the program behavior.

\paragraph{Loop optimization in symbolic execution} has been a topic of considerable interest, with a variety of strategies and techniques proposed in the literature to tackle the notorious path explosion problem arising due to loops. One such technique is loop summarization, which involves creating a summary of the loop's effect on the program state. This technique has been applied successfully in symbolic execution tools like Java PathFinder and KLEE to handle loops more efficiently.
Another significant approach is the use of loop invariants, which are conditions that remain true for each iteration of a loop. The work by Sharma et al.~\cite{sharma2012interpolants} used loop invariants to limit the number of symbolic executions of the loop body. The DART tool~\cite{godefroid2005dart} leverages selective symbolic execution, which decides to execute certain parts of the code concretely and others symbolically, useful for loops where some iterations don't affect the outcome of the execution.
The technique of bounded model checking involves putting an upper limit on the number of loop iterations, a strategy used in CBMC~\cite{clarke2004tool}. However, this technique may miss behaviors that occur beyond the set limit. Under-approximate loop acceleration is another method for dealing with loops, which aims to compute the postt-state after a certain number of iterations\cite{bardin2003fast}, but may lead to under-approximation of all possible behaviors.
While these methods provide different approaches to handle loops in symbolic execution, each comes with its trade-offs regarding completeness and precision. The selection of the appropriate loop handling strategy largely depends on the specific nature of the program and the computational resources available.

\subsection{Research Gap}
As discussed in the related work section~\ref{sec:RelWork}, numerous strategies have been proposed to enhance symbolic execution and broaden its capacity to solve larger problems~\cite{baldoni2018survey}. From our perspective, compositional approaches are especially intriguing. Current methodologies primarily focus on dissecting computer programs by function and methods, and calculating pre- and post-conditions. Godefroid~\cite{godefroid2007compositional} specifies preconditions as constraints in the memory inputs on the functions and postconditions as all the changes executed by the function during execution. For a function taking a parameter $x$, which returns $1$ if $x > 0$ and otherwise $0$, a summary would resemble: $x > 0 \land ret = 1 \lor x <=0 \land ret = 0$.

Most extant approaches for compositional symbolic execution focus on source code such as \texttt{C} or \texttt{Javascript}. The summaries mainly revolve around the transition from inputs (preconditions) to return values (postconditions). Godefroid~\cite{godefroid2007compositional}, however, also considered side effects by tracking memory writes. Memory corruption in the heap is a challenging issue~\cite{tu2023boosting}, in part because it depends on the heap implementation and symbolic heap simulation. In all cases, bugs such as \emph{Use-After-Free} may not be detectable, as the heap layout can vary due to factors like the operating system, installed memory, heap implementation version, etc. While certain patterns like duplicated free or use after free can be detected using the path constraints of symbolic execution, the full path may not be available when decomposing the computer program into functions or other slices.

Our compositional approach focuses on the symbolic execution of binaries, as opposed to source code. Although symbolic execution of source code is typically more efficient, compiler optimization may introduce slightly different behaviors in the binaries. Evaluating constraints on the binaries ensures the correctness of the binary. When dissecting the computer program into functions, we extract side effects like heap operations for later detection of hazardous paths through the computer program with interprocedural analysis. Our approach is compatible with loop optimization technologies such as automatically-inferred loop invariants, loop summaries, and decomposition of nested loops~\cite{godefroid2005dart}. Moreover, our approach can be used to create loop summaries.

Our contributions include the design and implementation of a function summary-based symbolic execution system, which hinges on program decomposition (functions, loops). Our system is capable of analyzing binaries for bugs and compiling side effects about heap operations for subsequent detection of heap errors. This allows us to identify different types of memory corruption, as well as instances where mitigation mechanisms such as stack canaries have been triggered. In such cases, even if they are often not exploitable, a program error has occurred that should be rectified. While some compositional systems may have false negatives compared to linear symbolic execution~\cite{godefroid2007compositional}, we allow for the possibility of false positives, as some summaries may permit more successor states than necessary, ensuring we do not have false negatives. Future work could eliminate false positives by verifying the feasibility of the path using directed or backward symbolic execution~\cite{ma2011directed}. Our evaluation confirms the viability of our approach and the speed improvements over classic symbolic execution.

\section{Smart Symbolic Summaries}
\label{sec:SmartSummaries}

This section presents the core components of our work, namely smart symbolic summaries. Our contribution includes a system that decomposes binaries into smaller slices and individually executes these slices (section \ref{sec:SmartSymbolicFunctionSummaries}), a concept for decomposing nested loops for improved loop performance (section \ref{sec:DecomposingNestedLoops}), and the extraction of certain side effects like heap operations (section \ref{sec:SideEffectExtraction}).

\subsection{Smart Symbolic Function Summaries}
\label{sec:SmartSymbolicFunctionSummaries}

Smart Symbolic Function Summaries form the basic concept for decomposing a computer program for faster symbolic execution. A summary, denoted as $\Sigma$, is an abstraction of a function or any other slice of the program, consisting of a set of preconditions $A$, a set of postconditions $\Omega$, and a set of side effects $\Theta$. The preconditions $A$ are defined as any possible constraint on the inputs (parameters, global variables, memory regions, etc.) which influence the resulting state of the function. Usually, each precondition $\alpha$ leads to one or more specific resulting states $\sigma$ accompanied by side effects $\theta$. We define a summary as:
\begin{equation}
\Sigma ( A ) \Rightarrow \langle \Omega ; \Theta \rangle
\end{equation}
A summary is a set of preconditions where each precondition maps to one or more postconditions and one or more side effects: $\alpha_p \rightarrow \langle \sum \omega \subset \Omega ; \sum \theta \subset \Theta \rangle$. Thus, we can also define a summary as:
\begin{equation}
\Sigma = \sum_{p_0}^{p_n} \sigma ( \alpha_p ) \Rightarrow \langle \Omega ; \Theta \rangle
\end{equation}
Therefore, a summary can be considered a lookup table, where the symbolic execution engine seeks a matching precondition and retrieves the corresponding postconditions and side effects. This process is much faster than executing a function every time on its own. Another advantage is that summaries can be cached and reused, meaning the function only needs to be executed once.

\begin{figure}
\centering
\includegraphics[width=.8\textwidth]{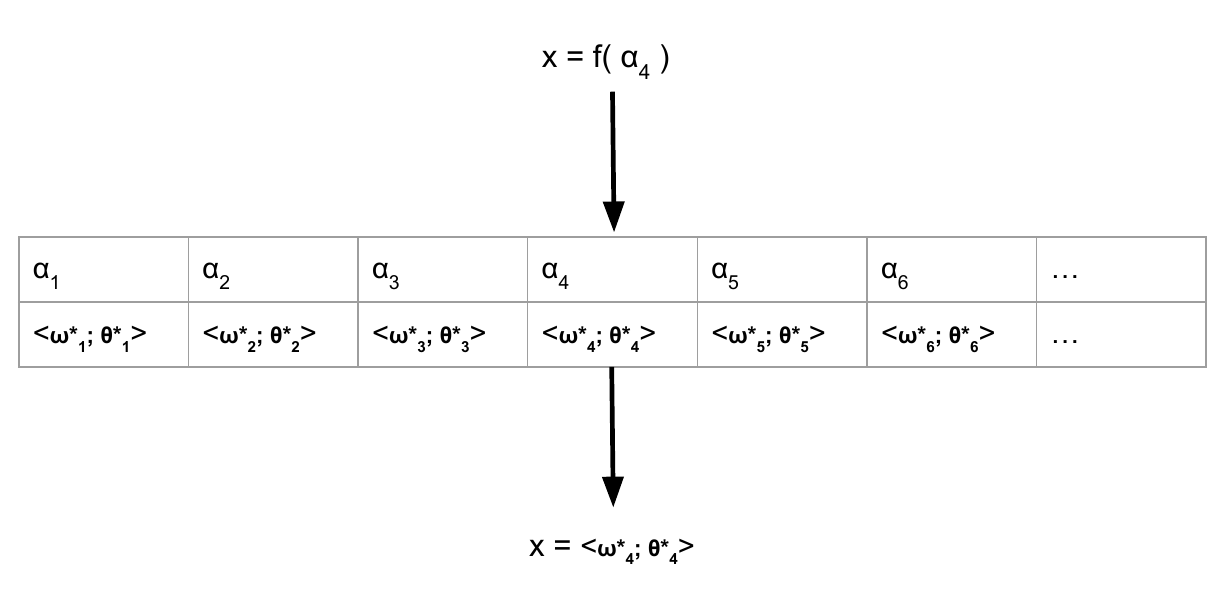}
\caption{Scheme of a function summary lookup table: instead of executing the function $f$, the summary $\sigma_4$ is applied, matching the precondition $\alpha_4$.}
\label{fig:func_summary_lookup}
\end{figure}

Applying symbolic execution and function summaries to binaries is slightly more complex than when executing the source code. This complexity arises not only from the differences among CPU architectures but also from the absence of information about function signatures. Extracting the type of a parameter from a binary function is challenging since memory regions can be interpreted in various ways. Moreover, we only have memory addresses for data storage and no variables, which makes executing a single function out of context challenging. However, it is crucial for the symbolic execution engine to set up the memory regions correctly to execute a function accurately. We tackle this issue by limiting the usage of constraints to memory regions where we can correctly understand the type. For other memory regions, we apply unconstrained symbolic values. This approach increases the possible search space and creates states which might not be reachable in the actual computer program. However, it ensures that we do not overlook any issues, at worst, only producing false positives.

The number of parameters, their types, and the return type can be difficult to extract from a binary, especially for architectures where parameters are passed by registers (for example, AMD64). The only way to analyze the number of parameters is to trace the data flow in the binary and check which registers are set before the function call. However, for binaries not compiled with default calling conventions, the recovery of the function call may not be possible at all.

For the $AMD64$ architecture, we can define a precondition as:
\begin{equation}
A = \langle |rdi|, |rsi|, |rdx|, |rcx|, |r8|, |r9|, \langle|rsp|\rangle \rangle
\end{equation}
Here, $|register|$ stands for the value or constraint on the register. The first six parameters are passed via registers, and additional parameters are pushed onto the stack. Not all registers are required for every function call. Registers used as parameters can be used as concrete or constrained symbolic values, while registers not used as parameters for a specific function call stay purely symbolic. Depending on the program state of a real execution, any value could be written in these registers, and they should be overwritten by the function before use.

It is also possible that the function depends on the state of global variables. The handling of global variables is complex for function summaries, since the state of global variables can depend on any previous operation in the computer program. Therefore, we try to add global variables to the side effects if possible and decide on them based on the recombination of individual slices/functions.

The primary post condition is the return value. Newly allocated or freed memory, as well as changed memory regions by \emph{call-by-reference} parameters, are counted as side effects. This makes the post conditions for $AMD64$ as follows:
\begin{equation}
\Omega = \langle |rax| \rangle
\end{equation}

For the $AMD64$ architecture, a function summary can be defined as:
\begin{equation}
\Sigma ( \langle |rdi|, |rsi|, |rdx|, |rcx|, |r8|, |r9|, \langle|rsp|\rangle \rangle ) \Rightarrow \langle |rax|; \Theta \rangle
\end{equation}

\subsection{Function Summary Execution Algorithm}
\label{sec:Function Summary Execution Algorithm}
After we defined what a smart symbolic summary is, we need to automatically create them. Therefore, we use an algorithm which is based on the SMART algorithm by Godefroid~\cite{godefroid2007compositional}, which is a top down approach. 
The concept of our algorithm is to execute the lowest functions in a call graph first, which do not call any other function or only call system functions. 
We start by creating summaries for these functions and then continue moving up in the function call tree, while applying previously created summaries. 

To create a function summary of a function $f$ we symbolically execute $f$ without applying any constraints on the parameter. The symbolic execution engine will find any possible path through $f$ and create all possible resulting states (postcondition) $\Omega$. 
For each resulting state $\omega_i$, we trace back under which precondition $\alpha_i$ the resulting state can be reached. Afterwards we create an entry for the function $f$ in our function lookup table (see figure \ref{fig:func_summary_lookup}): $\alpha_i \Rightarrow \omega_i$. After creating an entry for each possible resulting state, the summary of the function was created. 

For the $AMD64$ architecture it means we create a summary by setting all registers and if required the top of the stack to symbolic and execute the function. Therefore, it is important to consider that a parameter can also be \emph{call-by-reference} instead of \emph{call-by-value}, which requires additional memory setup. Afterwards, for each possible resulting state the $rax$ register is checked. Depending on the return type either nothing, a value, a symbolic value or a memory-reference is extracted. If a memory-reference is returned, the corresponding memory areas are important. Afterwards, we use the SMT-solver to find for each resulting state the constraints on the preconditions of the registers to create the summary.

\begin{figure}
    \centering
    \includegraphics[width=0.8\textwidth]{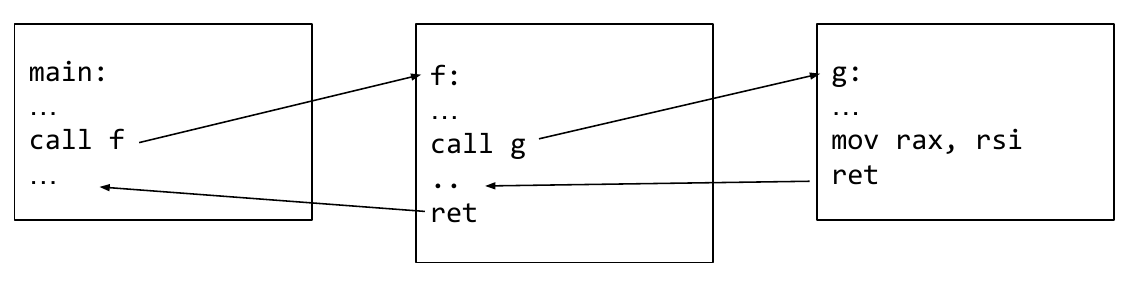}
    \caption{Callgraph: Main calls a function $f$. $f$ calls a function $g$. $g$ loads a return value in $rax$.}
    \label{fig:CallGraph_1}
\end{figure}

\paragraph{To traverse symbolically through a computer program} and create all required summaries as well as to analyze the relations between the functions we use a top-down approach.
We start at the main function. For now we assume command-line parameter are either specified by the user or not available. User-specified command-line parameters can be symbolic, but their length and number must be specified. 
Next, the main function is executed and we follow each possible path either by a Depth-First-Search (DFS) or by a Breadth-first search (BFS) until we reach a state with a function call $f$. At this point the execution of main is stopped and $f$ is executed individually. If $f$ calls another function $g$, the function $g$ is executed before the execution of $f$ is completed. After the execution of $g$ is finished, the summary of $g$ is applied to $f$, while the preconditions for executing $g$ are taken from the state of $f$, where $g$ is called. After applying the summary of $g$ to $f$, the execution of $f$ will be completed and the summary of $f$ will be applied to main accordingly. The corresponding call-graph is shown in figure \ref{fig:CallGraph_1}.

\begin{algorithm}[h!]
\caption{Symbolic Execution with Smart Function Summaries}
\label{alg:SymbolicExecutionwith smart function summaries}
\begin{algorithmic}[1]
\State FunctionSummaries: Dict $\gets$ init
\Procedure{SymbolicExecuteFunction}{FunctionName}
\State init registers used for parameter with unconstrained symbolic value
\State states $\gets$ InitState(FunctionName)
\While {states $\neq$ empty}
    \For{state $in$ states.CurrentStates}
        \State nextInst $\gets$ state.NextInstruction
        \If{nextInst $is\ a$ function call}
            \If{nextInst.FunctionName $is\ not\ in$ FunctionSummaries}
                \State SymbolicExecuteFunction(nextInst.FunctionName)
            \EndIf
            \State preconditions $\gets$ GetPreconditions(nextInst.FunctionName)
            \State ApplySummary(nextInst.FunctionName, preconditions)
        \Else
            \State ExecuteInstruction($nextInst$)
            \State CheckForBug(state)
        \EndIf
    \EndFor
\EndWhile
\EndProcedure
\State SymbolicExecuteFunction("main")
\end{algorithmic}
\end{algorithm}

In Algorithm \ref{alg:SymbolicExecutionwith smart function summaries} a pseudo-code implementation of our symbolic execution algorithm can be found. The algorithm assumes that there are no recursive function calls. To handle recursive calls it would be required to trace the call-stack, to figure out if the function was already called before. Since by definition a recursive function cannot be executed by function summaries at this point we assume the first recursive call returns a completely unconstrained value. 
The function $GetPreconditions$ extracts the required preconditions depending on the function named in $nextInst.FunctionName$. The function $ApplySummary$ applies the postconditions derived by the given preconditions (return value, etc) to the current function by adding the resulting states as to the Dict $states$. 
The function $InitState$ generates the required initial state from which the symbolic execution of the function will start.  
In case, a value cannot be determined, we assume the value to be unconstrained to ensure complete execution. 

Every time a instruction is executed we check if the state of the computer program fulfills any of the \emph{is-a-bug} conditions we defined. If a bug is detected, we log the bug and continue the execution. The simplest way to detect a critical bug (vulnerability) using symbolic execution is to check, whether the instruction pointer depends on user input or not. If it is possible to find a state where the instruction pointer depends on the input, any user could use this to redirect the control flow of the program, which would be a classic memory corruption vulnerability. Thereby, stack and heap overflows are already considered when checking if the instruction pointer could have been manipulated. Furthermore, we also check if the computer program ends up in certain states such as stack smashing, which occurs during a stack buffer overflow with an active stack canary. 

\subsection{Decomposing Nested Loops}
\label{sec:DecomposingNestedLoops}
Loops are a nightmare for symbolic execution, since they create a lot of states and therefore they massively slow symbolic execution down. While there are many 
solution approaches such as limiting the number of iterations, there is no \emph{good} solution to handle loops. Here, we make a suggestion on how loops can be handled, however, non of them will be a final solution for the problem but in certain situation the approaches will bring improvements. 

One of the largest problems with loops are nested loops, since nesting loops lead to polynomial runtime. For example, a simple loop with 10 iterations and an if-else statement is not a problem to be solved by a symbolic solver. However, if we have two nested loops, with both loops having 10 iterations each, we are already at 100 overall iterations, which brings us pretty close to the point where it becomes hard for symbolic execution. Therefore, instead of executing the inner loop every time the outer loop does one iteration, it makes sense to create a summary of the inner loop. This is possible, since every computer program with a nested loop can be rewritten into a computer program where the inner loop is in a separate function. 
For example, the computer program shown:
{\footnotesize
\begin{lstlisting}[language=C]
    for (i = 0; i < 10; ++i){
        for (j = 0; j < 10; ++j){
            //do something with i and j
        }
    }
\end{lstlisting}
}
can be rewritten to the following: 
{\footnotesize
\begin{lstlisting}[language=C]
    for (i = 0; i < 10; ++i){
        innerloop(i)
    }

    int innerloop(int i){
        for (j = 0; j < 10; ++j){
            //do something with i and j
        }
    }   
\end{lstlisting}
}
When symbolically executing the $innerloop$ function, a function summary is created, which can be applied for every iteration of the outer loop, reducing the number of states for the symbolic execution and the SMT solver. 
Together with a loop limiter, which limits the maximal number of iterations in a loop, this technique is very efficient to be able to solve larger problems and achieve a better coverage using symbolic execution. 
It is possible to detect loops in binary code by looking for jump instructions which are pointing to a previous location in the same function or by analyzing the control flow graph. Both methods can also be used to detect inner loops.
These methods are not limited to a single inner loop but can be applied to any number of nested loops and reduces the runtime from polynomial ($x^n$) to linear ($n\cdot) x$ for $n$ nested loops. However, this concept still cannot solve any loop, since the end of the loop may depend on user input which can make a loop infinite for a symbolic execution system or the number of iterations for a single loop is just too high to be symbolically solved. 

Therefore, we are looking into an approach for handling loops better based on bounded iterations. Since we decompose loops into smaller pieces it is possible to increase number of loop iterations in the loop limiter iteratively. As long as neither a time nor a memory constraint is reached by the loop limiter, the maximal number of iterations is slightly increase, so that the highest possible number of iterations for a given case is solved. 

\subsection{Side Effect Extraction}
\label{sec:SideEffectExtraction}

The function parameter $A$ and the return value $\Omega$ are not sufficient to create an effective function summary in a non-functional programming language. Therefore, we need to consider side effects $\Theta$ to execute functions as close to reality as possible and to avoid overlooking any bugs in the code. The focus is for now on two side effects: (1) heap memory operations and (2) call-by-reference parameter/global variables. An individual side effect $\theta$ is always a tuple of the operation and the corresponding memory address: $\theta = \langle Op, Mem \rangle$ and depends on a precondition $\alpha$.

For heap operations, we trace all $malloc$ and $free$ operations as well as all memory accesses on addresses that fall into the heap memory region. Thus, if a function executes a $malloc$ operation, a malloc will be added to the set of side effects $\Theta$. If the same memory address will be freed, the side effect 'free' will be added as a new side effect. This way we can later use the SMT solver to try to find a path through the computer program where the same memory address is freed twice or where a freed memory address is used. We add this check to our $CheckForBug$ function (see algorithm \ref{alg:SymbolicExecutionwith smart function summaries}). For every instruction executed by the symbolic execution engine, the $CheckForBug$ function verifies if there was a duplicated free or a UAF on the path.

For \emph{call-by-reference parameter}, the post-condition of the memory location at the end of the function is stored as a function of all preconditions $A$, and applied by the function $ApplySummary$ (see algorithm \ref{alg:SymbolicExecutionwith smart function summaries}) similar to all other post-conditions. The difficulty here is not to track or to apply the changes on the parameter but to understand correctly that we have a call-by-reference parameter for a binary program, since an address and an integer are difficult to distinguish. To understand if a parameter is a reference/pointer or an integer, the first access to the parameter is traced. If the parameter access is by a $lea$ instruction (load effective address in $AMD64$ architecture), or an array or memory access is detected, we assume the parameter is a reference/pointer and we trace the memory changes. If the parameter is changed by a $mov$ operation as the first operation (is overwritten), we can safely assume the parameter does not create a side-effect. If the first operation is an arithmetic operation, we cannot determine from this operation whether it is a reference/pointer or an integer, therefore we check subsequent instructions. If the parameter is copied into a different register, it is required to trace both registers until we can decide on the parameter or one of the registers is overwritten.

Global Variables can be handled similarly to call-by-reference parameters. However, we don't need to trace if we have an integer or pointer, since the memory region is neither stack nor heap. However, it needs to be considered that the global variable could be a pointer, which is allocated during the program execution. For the global variables, the postcondition after executing the function depending on the given precondition is stored and can be applied with the function $ApplySummary$.

\section{Evaluation}
\label{sec:Eval}
In this section we will present the current status of our implementation as well as the the evaluation we conducted. For transparancy we publish our source code as well as our test-set. The source code of our implementation can be found here: \url{https://github.com/FHNW-Digital-Trust/SwissSec}. The code is under development and may not be stable for all cases. For our tests we used the commit id \emph{18a9301}.
The test-set we used for our evaluation can be found here: \url{https://github.com/FHNW-Digital-Trust/TestSet} (commit \emph{2616df7}).

\subsection{Implementation}
Our implementation is based on \emph{angr}~\cite{shoshitaishvili2016state}, which is a versatile multi-architecture binary analysis toolkit, offering both static and dynamic analysis capabilities. One of its powerful features is symbolic execution, a technique used to analyze a program to determine what inputs cause each part of a program to execute. As solver \emph{angr} uses \emph{Z3}~\cite{de2008z3}, which is a very efficient SMT solver. Our implementation currently covers the function summary execution algorithm from section \ref{sec:Function Summary Execution Algorithm} as well as the possibility to trace operations.
However, the precondition extraction works only with a limited number of parameter. Solving for duplicated free or UAF is not yet implemented and memory tracing is unstable. Nevertheless, we want to present our current work as prove, that efficient symbolic execution is also possible on binary level to find software weaknesses. 
The name of our implementation is \emph{SwissSec}.

\subsection{Results}
Our evaluation setup is based on our implementation. We run our test programs on a computer with an AMD Ryzen 7950X with 128GB of memory on Ubuntu 22.04. We run each test-program 100 times and took the median value. As baseline for the comparison, we use the default strategy of \emph{angr}, since we did not find any other compositional symbolic execution implementation which supports the analysis of binaries. 

Our implementation aims to evaluate the speed of execution as well as how large the largest problem is that can be solved.
Our test set consists of 12 programs, each with varying levels of difficulty for symbolic execution. In each program a bug is hidden which needs to be found by the symbolic execution engine. 

\begin{figure}[h!]
    \centering
  \begin{tikzpicture}[scale=0.66]
    \begin{axis}[
        ybar=2pt,  
        bar width=.3cm,
        width=\textwidth,
        height=.5\textheight,
        legend style={at={(0.05,0.95)},anchor=north west},
        ylabel={Time in Seconds},
        symbolic x coords={
            Simple1,
            SimpleFunctionCalls,
            SimpleFunctionWithContraints,
            SimpleFunctionWithContraints2,
            SimpleLibTest,
            SimpleLoop,
            SimpleUAF,
            LoopAndFunctionCall,
            LoopsAndFunctionCalls,
            NestedLoopLevel1,
            NestedLoopLevel2
        },
        xtick=data,
        nodes near coords,
        nodes near coords align={vertical},
        x tick label style={rotate=45,anchor=east},
    ]
    \addplot coordinates {
        (Simple1,0.767246246)
        (SimpleFunctionCalls,2.301935911)
        (SimpleFunctionWithContraints,0.530171633)
        (SimpleFunctionWithContraints2,0.549474239)
        (SimpleLibTest,0.247897387)
        (SimpleLoop,1.165613174)
        (SimpleUAF,0.759821653)
        (LoopAndFunctionCall,1.433349609)
        (LoopsAndFunctionCalls,0.419448853)
        (NestedLoopLevel1,0.959580898)
        (NestedLoopLevel2,1.014062166)
    };
    \addplot coordinates {
        (Simple1,1.721290827)
        (SimpleFunctionCalls,1.912759304)
        (SimpleFunctionWithContraints,1.51222682)
        (SimpleFunctionWithContraints2,1.526368618)
        (SimpleLibTest,1.014615297)
        (SimpleLoop,2.146978855)
        (SimpleUAF,1.742531538)
        (LoopAndFunctionCall,6.862039328)
        (LoopsAndFunctionCalls,1.588543892)
        (NestedLoopLevel1,1.490966797)
        (NestedLoopLevel2,5.247635603)
    };
    \legend{SwissSec, Angr}
    \end{axis}
\end{tikzpicture}
    \caption{Results of the comparison between \emph{SwissSec} and \emph{angr} (baseline) on our test-set.}
    \label{fig:res_testset}
\end{figure}
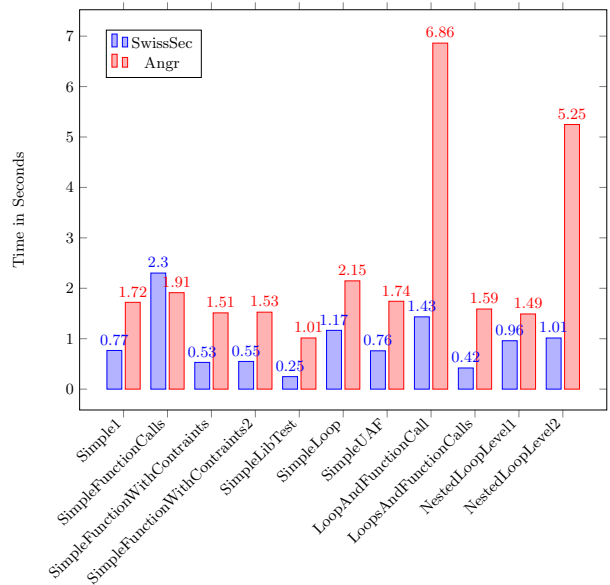

Figure \ref{fig:res_testset} shows a comparison of the performance of \emph{SwissSec} and \emph{angr} on our test-set. We especially notice that the performance for loops where additional operation such as function calls or inner loops in function calls are used, the performance of our tool is far better. Given the principles of our approach, this is exactly what we expected. For the test $SimpleFunctionCalls$ we can see, that the overhead of splitting into functions can also reduce the speed due to the overhead in certain cases. Usually, this happens with very small program, which are executed fast in any case. 

\begin{figure}
    \centering
    \begin{tikzpicture}[scale=0.75]
        \begin{axis}[
            xlabel={Nested Loops},
            ylabel={Time in Seconds},
            legend pos=north west,
            ymajorgrids=true,
            grid style=dashed,
        ]
        \addplot[color=blue,mark=square,]
        coordinates {
            (1,0.959580898)
            (2,1.014062166)
            (3,1.121983051)
            (4,1.20545578)
            (5,1.357113123)
        };
        \addplot[color=red,mark=*,]
        coordinates {
            (1,1.490966797)
            (2,5.247635603)
            (3,79.1174109)
            (4,1632.2887303829193)
            (5,nan)
        };
        \legend{SwissSec, Angr}
        \end{axis}
    \end{tikzpicture}
    \caption{Results of the comparison between \emph{SwissSec} and \emph{angr} (baseline) for tests on nested loops.}
    \label{fig:testloops}
\end{figure}
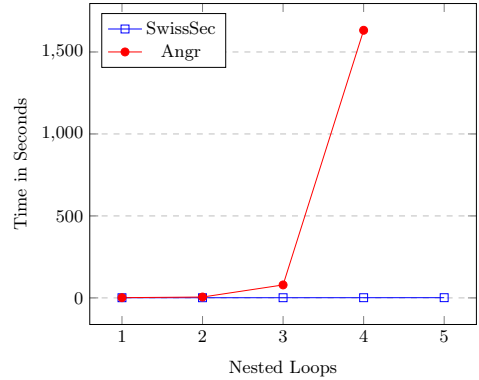

As second part of our evaluation, we want to check the scaling of loops when adding additional nested loops and function calls. We expect that, since SwissSec decomposes the computer program, it will scale linearly, in contrast to Angr, which we expect to scale with the complexity of the loops (polynomially). The result in figure \ref{fig:testloops} confirms this. For the test case with 5 nested loops, Angr was unable to solve the problem within the 60-minute time frame we allotted for the test. Note: the polynomial increase of states for the nested loop leads to an exponential increase in time for the symbolic execution due to the NP-hardness of the solver.

\section{Conclusion \& Future Work}
\label{sec:Conclusion}

In this paper, we introduce a divide-and-conquer approach to the symbolic execution of binaries. Our aim is to minimize the problem size for the SMT solver and tackle the path explosion problem. The crux of our strategy is to execute individual slices of a computer program—like functions—and subsequently integrate the results. When implementing this approach for binaries, we encountered several challenges not typically faced by approaches focusing on symbolic execution of source code. For instance, extracting a function's prototype from binary code isn't always straightforward.
Alongside this simple decomposition based on function code, we devised methods to slice nested loops, thereby reducing the computational complexity of symbolically executing such structures. Additionally, we demonstrated how to extract side effects from functions, enabling us not only to detect common memory corruption bugs, such as buffer overflows, but also more complex issues like Use-After-Free scenarios.

Our divide-and-conquer method is implemented in a tool called SwissSec, which is built upon the foundation of Angr. Our evaluation highlights the advantages of our approach over classic symbolic execution with Angr, particularly for binaries that contain loops and function calls.

Looking ahead, our future work will involve developing function summaries that can extract and evaluate more side effects, thereby enhancing our ability to detect complex bugs and even aiding in the evaluation of business logic. Our intention is to extract the side effects as features from the function summaries and incorporate them into a model checker that verifies the business logic. The primary benefit of this approach is that the modeled functionality is guaranteed to match the actual program, since it is directly extracted from it.

Other points for future development include the integration of more bounded checking systems. These systems will iteratively add complexity, such as automatically estimating command-line parameters and further improving loop coverage.

\newpage
%
%
%
 \bibliographystyle{splncs04}
 \bibliography{mybibliography}

\begin{thebibliography}{10}
\providecommand{\url}[1]{\texttt{#1}}
\providecommand{\urlprefix}{URL }
\providecommand{\doi}[1]{https://doi.org/#1}

\bibitem{anand2008demand}
Anand, S., Godefroid, P., Tillmann, N.: Demand-driven compositional symbolic
  execution. In: Tools and Algorithms for the Construction and Analysis of
  Systems: 14th International Conference, TACAS 2008, Held as Part of the Joint
  European Conferences on Theory and Practice of Software, ETAPS 2008,
  Budapest, Hungary, March 29-April 6, 2008. Proceedings 14. pp. 367--381.
  Springer (2008)

\bibitem{avgerinos2014enhancing}
Avgerinos, T., Rebert, A., Cha, S.K., Brumley, D.: Enhancing symbolic execution
  with veritesting. In: Proceedings of the 36th International Conference on
  Software Engineering. pp. 1083--1094 (2014)

\bibitem{baldoni2018survey}
Baldoni, R., Coppa, E., D’elia, D.C., Demetrescu, C., Finocchi, I.: A survey
  of symbolic execution techniques. ACM Computing Surveys (CSUR)
  \textbf{51}(3),  1--39 (2018)

\bibitem{bardin2003fast}
Bardin, S., Finkel, A., Leroux, J., Petrucci, L.: Fast: Fast acceleration of
  symbolic transition systems. In: Computer Aided Verification: 15th
  International Conference, CAV 2003, Boulder, CO, USA, July 8-12, 2003.
  Proceedings 15. pp. 118--121. Springer (2003)

\bibitem{burnim2008heuristics}
Burnim, J., Sen, K.: Heuristics for scalable dynamic test generation. In: 2008
  23rd IEEE/ACM International Conference on Automated Software Engineering. pp.
  443--446. IEEE (2008)

\bibitem{cadar2013symbolic}
Cadar, C., Sen, K.: Symbolic execution for software testing: three decades
  later. Communications of the ACM  \textbf{56}(2),  82--90 (2013)

\bibitem{clarke2004tool}
Clarke, E., Kroening, D., Lerda, F.: A tool for checking ansi-c programs. In:
  Tools and Algorithms for the Construction and Analysis of Systems: 10th
  International Conference, TACAS 2004, Held as Part of the Joint European
  Conferences on Theory and Practice of Software, ETAPS 2004, Barcelona, Spain,
  March 29-April 2, 2004. Proceedings 10. pp. 168--176. Springer (2004)

\bibitem{de2008z3}
De~Moura, L., Bj{\o}rner, N.: Z3: An efficient smt solver. In: International
  conference on Tools and Algorithms for the Construction and Analysis of
  Systems. pp. 337--340. Springer (2008)

\bibitem{geldenhuys2012probabilistic}
Geldenhuys, J., Dwyer, M.B., Visser, W.: Probabilistic symbolic execution. In:
  Proceedings of the 2012 International Symposium on Software Testing and
  Analysis. pp. 166--176 (2012)

\bibitem{godefroid2007compositional}
Godefroid, P.: Compositional dynamic test generation. In: Proceedings of the
  34th annual ACM SIGPLAN-SIGACT symposium on Principles of programming
  languages. pp. 47--54 (2007)

\bibitem{godefroid2005dart}
Godefroid, P., Klarlund, N., Sen, K.: Dart: Directed automated random testing.
  In: Proceedings of the 2005 ACM SIGPLAN conference on Programming language
  design and implementation. pp. 213--223 (2005)

\bibitem{irlbeck2015deconstructing}
Irlbeck, M., et~al.: Deconstructing dynamic symbolic execution. Dependable
  Software Systems Engineering  \textbf{40}(2015), ~26 (2015)

\bibitem{king1976symbolic}
King, J.C.: Symbolic execution and program testing. Communications of the ACM
  \textbf{19}(7),  385--394 (1976)

\bibitem{lin2015compositional}
Lin, Y., Miller, T., S{\o}ndergaard, H.: Compositional symbolic execution using
  fine-grained summaries. In: 2015 24th Australasian Software Engineering
  Conference. pp. 213--222. IEEE (2015)

\bibitem{ma2011directed}
Ma, K.K., Yit~Phang, K., Foster, J.S., Hicks, M.: Directed symbolic execution.
  In: Static Analysis: 18th International Symposium, SAS 2011, Venice, Italy,
  September 14-16, 2011. Proceedings 18. pp. 95--111. Springer (2011)

\bibitem{ramos2023toward}
Ramos, F., Sabino, N., Ad{\~a}o, P., Naumann, D.A., Fragoso~Santos, J.: Toward
  tool-independent summaries for symbolic execution. In: 37th European
  Conference on Object-Oriented Programming (ECOOP 2023). Schloss
  Dagstuhl-Leibniz-Zentrum f{\"u}r Informatik (2023)

\bibitem{sen2005cute}
Sen, K., Marinov, D., Agha, G.: Cute: A concolic unit testing engine for c. ACM
  SIGSOFT Software Engineering Notes  \textbf{30}(5),  263--272 (2005)

\bibitem{sharma2012interpolants}
Sharma, R., Nori, A.V., Aiken, A.: Interpolants as classifiers. In:
  International Conference on Computer Aided Verification. pp. 71--87. Springer
  (2012)

\bibitem{shoshitaishvili2016state}
Shoshitaishvili, Y., Wang, R., Salls, C., Stephens, N., Polino, M., Dutcher,
  A., Grosen, J., Feng, S., Hauser, C., Kruegel, C., Vigna, G.: {SoK: (State
  of) The Art of War: Offensive Techniques in Binary Analysis}. In: IEEE
  Symposium on Security and Privacy (2016)

\bibitem{trabish2018chopped}
Trabish, D., Mattavelli, A., Rinetzky, N., Cadar, C.: Chopped symbolic
  execution. In: Proceedings of the 40th International Conference on Software
  Engineering. pp. 350--360 (2018)

\bibitem{tu2023boosting}
Tu, H.: Boosting symbolic execution for heap-based vulnerability detection and
  exploit generation. In: 2023 IEEE/ACM 45th International Conference on
  Software Engineering: Companion Proceedings (ICSE-Companion). pp. 218--220.
  IEEE (2023)

\bibitem{zhu2022fuzzing}
Zhu, X., Wen, S., Camtepe, S., Xiang, Y.: Fuzzing: a survey for roadmap. ACM
  Computing Surveys (CSUR)  \textbf{54}(11s),  1--36 (2022)

\end{thebibliography}
\appendix

\newpage
\section*{Appendix}
\subsection*{Measurment Data}

Average value (out of 100 runs) of the measurement data for \ref{fig:res_testset} can be found in table \ref{tab:Mesuurment data 1}.
\begin{table}[ht]
\centering
\begin{tabularx}{\textwidth}{|X|c|c|}
\hline
Testprogram & SwissSec & Angr \\
\hline
Simple1 & 0.767246246 & 1.721290827 \\
\hline
SimpleFunctionCalls & 2.301935911 & 1.912759304 \\
\hline
SimpleFunctionWithContraints & 0.530171633 & 1.51222682 \\
\hline
SimpleFunctionWithContraints2 & 0.549474239 & 1.526368618 \\
\hline
SimpleLibTest & 0.247897387 & 1.014615297 \\
\hline
SimpleLoop & 1.165613174 & 2.146978855 \\
\hline
SimpleUAF & 0.759821653 & 1.742531538 \\
\hline
LoopAndFunctionCall & 1.433349609 & 6.862039328 \\
\hline
LoopsAndFunctionCalls & 0.419448853 & 1.588543892 \\
\hline
NestedLoopLevel1 & 0.959580898 & 1.490966797 \\
\hline
NestedLoopLevel2 & 1.014062166 & 5.247635603 \\
\hline
\end{tabularx}
\caption{Measurement Data for figure \ref{fig:res_testset}.}
\label{tab:Mesuurment data 1}
\end{table}

Average value (out of 100 runs) of the measurement data for figure \ref{fig:testloops} can be found in table \ref{tab:nested_loops}.
\begin{table}[ht]
\centering
\begin{tabularx}{\textwidth}{|X|c|c|}
\hline
Test Case & SwissSec & Angr \\
\hline
1 Nested Loop & 0.959580898 & 1.490966797 \\
\hline
2 Nested Loops & 1.014062166 & 5.247635603 \\
\hline
3 Nested Loops & 1.121983051 & 79.1174109 \\
\hline
4 Nested Loops & 1.20545578 & 1632.2887303829193 \\
\hline
5 Nested Loops & 1.357113123 & unsolved withing 60min \\
\hline
\end{tabularx}
\caption{Measurement Data for Nested Loops (figure \ref{fig:testloops}).}
\label{tab:nested_loops}
\end{table}

\subsection*{Angr-Baseline Script}
{\tiny
\begin{lstlisting}[language=python]
import sys
import angr, claripy
import time
binary = sys.argv[1]
t = time.time()
p = angr.Project(binary)
state = p.factory.entry_state()
state.options.add(angr.sim_options.SYMBOL_FILL_UNCONSTRAINED_REGISTERS)
state.options.add(angr.sim_options.SYMBOL_FILL_UNCONSTRAINED_MEMORY) 
state.register_plugin("heap", angr.state_plugins.heap.heap_ptmalloc.SimHeapPTMalloc())
simgr = p.factory.simgr(state, save_unconstrained=True)
simgr.run(until=lambda sm: len(sm.unconstrained) > 0)
et = time.time() - t
print ("Time", et)
\end{lstlisting}
}

\end{document}